\documentclass[aps,twocolumn,prb,amssymb,floatfix,10pt]{revtex4-1}

\usepackage{graphicx}
\usepackage{subfigure}

\begin{document}

\title{A model for the dynamics and internal structure of planar doping fronts in organic semiconductors}

\author{M. Modestov}
\affiliation{Department of Physics, Ume{\aa} University, SE--901 87 Ume{\aa}, Sweden}

\author{V. Bychkov}
\affiliation{Department of Physics, Ume{\aa} University, SE--901 87 Ume{\aa}, Sweden}

\author{D. Valiev}
\affiliation{Department of Physics, Ume{\aa} University, SE--901 87 Ume{\aa}, Sweden}

\author{M. Marklund}
\email[E-mail address: ]{mattias.marklund@physics.umu.se}
\affiliation{Department of Physics, Ume{\aa} University, SE--901 87 Ume{\aa}, Sweden}

\begin{abstract}
The dynamics and internal structure of doping fronts in organic
semiconductors are investigated theoretically using an extended
drift-diffusion model for ions, electrons and holes. The model  also
involves the injection barriers for electrons and holes in the
partially doped regions in the form of the Nernst equation, together with a
strong dependence of the electron and hole mobility on
concentrations. Closed expressions for the front velocities and  the ion concentrations in the doped regions are obtained. The analytical theory is
employed to describe the acceleration of the p- and n-fronts towards
each other. The analytical results show very good agreement with the
experimental data. Furthermore, it is shown that the internal structure of the doping
fronts is determined by the diffusion and mobility processes.
The asymptotic behavior of the concentrations and the electric field
is studied analytically inside the doping
fronts. The numerical solution for the front structure confirms the
most important predictions of the analytical theory: a sharp
head of the front in the undoped region, a smooth relaxation
tail in the doped region, and a plateau at the critical point of
transition from doped to undoped regions.
\end{abstract}

\maketitle

\section{Introduction}

Organic semiconductors (OSCs) demonstrate a number of interesting
properties, which distinguish them from crystalline inorganic
semiconductors \cite{Malliaras,Sirringhaus,Heeger}. Among the most intriguing features is the possibility of electrochemical doping by means of reversible redox reactions, demonstrated in a number of materials \cite{Forrest,Leger}. The electrochemical doping transforms the OSCs from essentially insulating state to a metallic-like state with the conductivity increased by many orders of magnitude \cite{Chiang}. This transformation is accompanied by a considerable change of the material properties including color, photoluminescence capability, volume, and surface energy \cite{Heeger,Leger,Sirringhaus}.
The electrochemical doping can be performed straightforward \textit{in situ}, when an OSC coated on a metal electrode is in contact with an electrolyte and is subjected to an appropriate electric potential. In the transformation process, electronic charges (electrons and holes) from the electrode are injected into the OSC and subsequently electrostatically compensated by an influx of respective ions from the electrolyte (cations and anions) \cite{Li,Pei-96,Matyba-08}.

The opportunity for tuning the electronic and optical properties of OSCs has triggered a number of studies at the fundamental as well as practical aspects of doping. These works in turn paved the way to numerous applications of this process in novel electronic and photonic devices \cite{Malliaras,Pei-95,Heeger,Sirringhaus,Coropceanu,Forrest,Leger,Bredas,Arkhipov}. A classical example of the electronic devices utilizing the electrochemical doping is polymer based light-emitting electrochemical cells (LECs) \cite{Pei-95,Pei-96,Pei-97,Sun,Matyba-09}. A LEC comprises an organic semiconductor in a form of conjugated fluorescent polymer, which is blended with a solid state electrolyte. The blend of the conjugated polymer and the electrolyte forms the active material of the LEC and is sandwiched between two electrodes.
When an electric potential applied between the electrodes exceeds the band gap potential of the conjugated polymer, then the doping transformation starts with injected holes and electrons forming the p- and n-type doped regions close to the respective electrodes \cite{Reenen}.
The doping process in planar LECs can be visualized  under ultra-violet illumination since doped OSCs display very high rate of photoluminescence quenching.
 The doped material in LEC is observed as dark regions quite distinct from the original undoped substance.
 By employing the ultra-violet visualization, it was demonstrated experimentally
 that doping transformation in OSC develops in the form of
two localized fronts of p- and n-type doping, which emerge at the electrodes and propagate towards each other \cite{Gao-04,Hu-06,Matyba-09,Reenen,Johansson,Robinson}.
When the  fronts meet, the two doped regions form a p-n junction, which emits a visible light. The purpose of
the present work is to provide a theoretical model for the front
propagation in LECs prior to the development of the p-n junction.

A number of interesting theoretical works has been devoted to
charge dynamics in LECs and other OSC devices
\cite{Smith,Manzanares,Lacroix,Miomandre,Wang-04,Wang-09,Johansson}.
In particular, the authors of Refs. \onlinecite{Smith,Manzanares} investigated a
stationary light-emitting p-n junction as a final state of the doping process in
LECs. At this stage of the process, the whole OSC is already converted to the state with high conductivity.
However, the non-steady problem of doping front dynamics and
structure prior to the formation of the p-n junction
is much more difficult to analyze, since it involves transition of the OCS from
the undoped weakly conducting state to the  metallic-like doped state within the front.
As
we show in the present paper, the problem includes not only
the electrodynamic issues of the OSC plasma motion, but also
some questions of thermodynamics and quantum mechanics, which are
still waiting for an answer.
The problem of front dynamics was addressed in
\onlinecite{Johansson,Robinson} from the empirical point of view of total
OSC conduction: the purpose was to analyze the
experimental data without investigating the complicated internal front
properties.
At the same time, considerable progress has been
achieved in the adjacent problem of front dynamics in one-electrode OSC devices, such as e.g. electrochemical sensors and actuators \cite{Lacroix,Wang-04,Wang-09}.
For example, the recent work by Wang et al. \cite{Wang-09}
provided discussion of all elements required for front formation in
the one-electrode devices.
The  model proposed in Ref. \onlinecite{Wang-09} included the Poisson
equation, diffusion and mobility (migration) of holes and ions
(cathions), taking into account possible nonlinear (non-Fickian)
character of the transport coefficients. The Nernst equation for
light particles, holes, in the region of high conductivity in that
case follows from the diffusion-mobility model under the condition
of quasi-equilibrium.
In a sense, the process studied in
Ref. \onlinecite{Wang-09} is just the opposite to the doping fronts propagating in LECs.
In
the doping process in LECs, holes and electrons are injected into
the active material by an externally applied electrical field,
which helps them overcoming a certain thermodynamic barrier, and dope the active material.
The ions, on the other hand, give way to the light charges, and compensate the excessive charge thus avoiding generation of strong
internal electric fields.
In the devices presented by Wang at al. \cite{Wang-09} charge motion
goes in the opposite direction: the cathions advance
together with the front, while holes retreat leaving the material.
With some caution, the fronts of electrochemical transformation in
LECs and in the one-electrode devices \cite{Wang-09} may be compared to uphill climb
and downhill glide, respectively.
Still, the physical understanding of front
dynamics in the one-electrode devices as presented in \cite{Wang-09} may be helpful in constructing
 models of the doping front propagation in LECs.
 In particular,
recent paper \cite{Modestov-10} demonstrated that the theoretical
description of  doping fronts in LECs requires not only the
common diffusion-mobility set of equations, but also  nonlinear
concentration-dependent transport coefficients for holes/electrons
and the thermodynamic injection
barrier for the light charges in the form of the Nernst  potential.

The present paper develops the ideas of Refs.
\cite{Wang-09,Modestov-10}. Using the theoretical model proposed in
\cite{Modestov-10}, we study dynamics and internal structure of the
doping fronts in LECs. We derive compact analytical formulas for the
front velocities depending on the electric field together with the
ion concentrations in the doped regions. On the basis of the
analytical theory we describe acceleration of the p- and n-fronts
approaching each other. The analytical results show very good agreement
with the experimental data. We show that the internal structure of the
doping fronts is determined by  diffusion and mobility processes.
We  study analytically the asymptotic behavior of the concentrations
and the electrical field inside  the doping
fronts. We also  solve numerically for the front structure,
and confirm the most important predictions of the analytical
theory: a sharp head of the front in the undoped region, a
smooth relaxation tail in the doped region, and a plateau at the
critical point of transition from doped to undoped regions.

The paper is organized as follows: In Sec. II we introduce the basic
equations of the model. In Sec. III we study properties of the doping fronts
considered as  surfaces of discontinuity; we find the front velocity and ion
concentrations in the doped regions behind the fronts and then describe front
acceleration in LECs. Section IV is devoted to the internal structure of the
fronts. We discuss characteristic length scales of the process, the
condition of quasi-neutrality, and the necessity of an injection barrier for
the system of equations. We investigate asymptotic analytical solutions to
the equations in the specific zones of the front and, finally, we numerically solve the
whole set of equations. The paper is concluded by a brief
summary in Sec. V.

\section{The mobility-diffusion approach}

In general, the dynamics of the ions is determined by equations of force balance
\begin{equation}
\label{eq1}
nm{\frac{{d{\rm {\bf v}}}}{{dt}}} = - qn\nabla \phi - k_{B} T\nabla n -
{\frac{{1}}{{\tau}} }nm{\rm {\bf v}},
\end{equation}
where $q = \pm e$ is charge of positive/negative ions, $m$
is ion mass, $n$ is concentration, $\phi $ is potential of an
electric field with ${\rm {\bf E}} = - \nabla \phi $, $k_{B} $ is
the Boltzmann constant and $T$ is the polymer temperature, which may be
taken constant and uniform. The last term in Eq. (\ref{eq1}) takes
the average contribution of collisions into account. In the case of
organic polymers, the collisions dominate over the inertia terms, which
allows the mobility-diffusion approach, giving the velocity according to
\begin{equation}
\label{eq2}
{\rm {\bf v}}_{\pm}  = \mp \mu _{\pm}  \nabla \phi - {\frac{{D_{\pm}
}}{{n_{\pm}} } }\nabla n_{\pm}  ,
\end{equation}
where the mobility is given by $\mu = e\tau / m$, labels $\pm $ correspond
to positive and negative ions, respectively, and the diffusion coefficient
$D$ is related to mobility using the Einstein relation $\mu = eD / k_{B} T$.
When substituting velocity from Eq. (\ref{eq2}) to the continuity equations
\begin{equation}
\label{eq3}
{\frac{{\partial n_{\pm}} } {{\partial t}}} + \nabla \cdot (n_{\pm}  {\rm
{\bf v}}_{\pm}  ) = 0,
\end{equation}
we arrive to the mobility-diffusion model for ion motion
according to
\begin{equation}
\label{eq4}
{\frac{{\partial n_{\pm}} } {{\partial t}}} - \nabla \cdot {\left[ {\pm \mu
_{\pm}  n_{\pm}  \nabla \phi + D_{\pm}  \nabla n_{\pm}}   \right]} = 0{\rm
.}
\end{equation}
The equations for electrons and holes may be presented in a similar way,
though in the case of light charges one has to take into account an
injection barrier $\phi _{N} $ for the electrons and holes in the
transition from the doped regions to the undoped ones
\begin{equation}
\label{eq5}
{\frac{{\partial n_{h}}} {{\partial t}}} - \nabla \cdot {\left[ {\mu _{h}
n_{h} \nabla (\phi - \phi _{N} ) + D_{h} \nabla n_{h}}  \right]} = 0,
\end{equation}
\begin{equation}
\label{eq6}
{\frac{{\partial n_{e}}} {{\partial t}}} - \nabla \cdot {\left[ { - \mu _{e}
n_{e} \nabla (\phi - \phi _{N} ) + D_{e} \nabla n_{e}}  \right]} = 0,
\end{equation}
where the labels ``h'' and ``e'' stand for holes and electrons, respectively.
The barrier originates from quantum and thermodynamic effects, and the
interpretation of $\phi _{N} $ in terms of an electrostatic force should
therefore be done with caution. We discuss the necessity of $\phi _{N} $ for
self-consistent description of the doping front structure and the particular
form of this term in the Section IV. Meanwhile we point out that this term
is non-zero only inside the doping fronts, and it should turn to zero in the
uniform undoped and doped material ahead of the fronts and behind the
fronts, respectively. The electric field obeys the Poisson equation
\begin{equation}
\label{eq7}
\nabla ^{2}\phi = (n_{ -}  + n_{e} - n_{ +}  - n_{h} )e / \varepsilon _{0}
.
\end{equation}
Still, in Section IV we demonstrate  that the condition of quasi-neutrality
\begin{equation}
\label{eq8}
n_{ -}  + n_{e} - n_{ +}  - n_{h} = 0
\end{equation}
holds with a very good accuracy in LECs. We stress that
condition (\ref{eq8}) \textit{does not} mean constant electric field everywhere. Instead, it implies
that even tiny local deviations from zero net charge lead to extremely large
electric fields.

\section{Properties of discontinuous doping fronts}

It was demonstrated experimentally
\cite{Matyba-09,Johansson,Robinson} that the doping process in OSCs happens in a form of two fronts propagating towards
each other as shown schematically in Fig. 1. A p-doping front
populates the semiconductor with holes, while the n-front makes the
semiconductor rich with electrons.
\begin{figure}
\includegraphics[width=3.5in,height=2.0in]{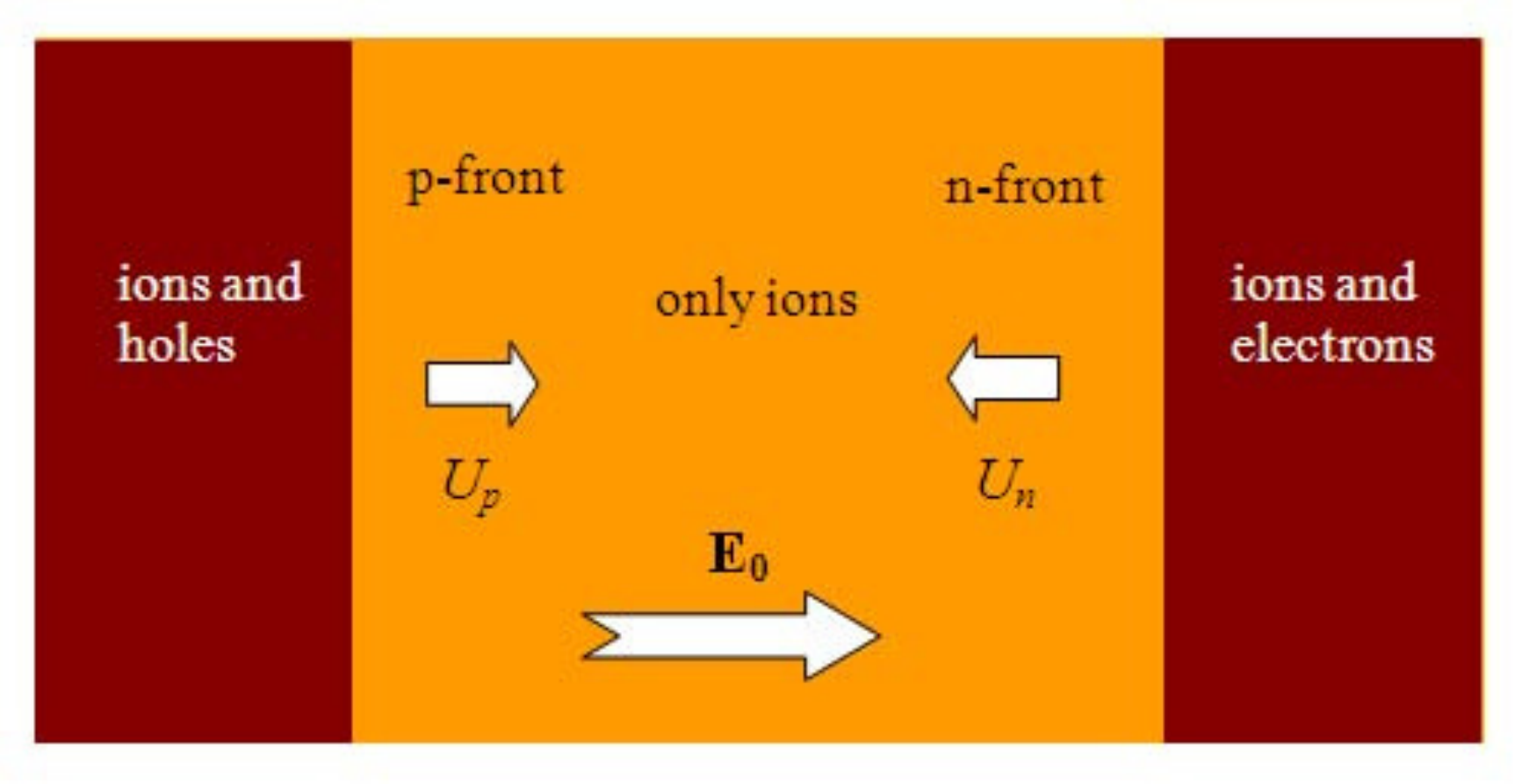}
\caption{Schematic of the doping process. }
\end{figure}
Both fronts propagate due to external electric field created by the
potential difference between the ends of the semiconductor film. In
the present section we consider planar p- and n-fronts as
propagating discontinuity surfaces, which transform the original
undoped semiconductor to a doped conducting material. To be
particular, we start our analysis with a stationary p-doping front propagating
 with velocity $U_{p} $ along the x-axis in a static and uniform
electric $E_{0} = const$, created in the undoped region by external
sources. Looking for the p-front solution in the form $\Psi = \Psi
(x - Ut)$ we reduce Eqs. (\ref{eq4}), (\ref{eq5}), (\ref{eq7}) to
\begin{equation}
\label{eq9}
{\frac{{d}}{{dx}}}\left[ { - n_{h} U_{p} + n_{h} \mu _{h} (E - E_{N} ) -
D_{h} {\frac{{dn_{h}}} {{dx}}}} \right] = 0,
\end{equation}
\begin{equation}
\label{eq10}
{\frac{{d}}{{dx}}}\left( { - n_{ -}  U_{p} - n_{ -}  \mu _{ -}  E - D_{ -}
{\frac{{dn_{ -}} } {{dx}}}} \right) = 0,
\end{equation}
\begin{equation}
\label{eq11}
{\frac{{d}}{{dx}}}\left( { - n_{ +}  U_{p} + n_{ +}  \mu _{ +}  E - D_{ +}
{\frac{{dn_{ +}} } {{dx}}}} \right) = 0,
\end{equation}
\begin{equation}
\label{eq12}
{\frac{{dE}}{{dx}}} = e(n_{ +}  - n_{ -}  + n_{h} ) / \varepsilon _{0} {\rm
,}
\end{equation}
where ${\rm {\bf E}}_{N} \equiv - \nabla \phi _{N} $ is
the effective electric field related to the quantum-thermodynamic
injection barrier. Still, in the limit of discontinuous doping
fronts considered in the present section, the barrier term does not
play any role, since it is zero in the uniform regions both ahead of
the front and behind it. Equations (\ref{eq9}) -- (\ref{eq12}) may
be integrated analytically across the front. We designate values
ahead of the front by the label ``0'' and values behind the front by
``1''. We also take into account that the hole concentration is zero
ahead of the front and it reaches some known finite value $n_{1h} $ behind
the front. The initial concentrations of positive and negative ions are equal
and known, $n_{0 -}  = n_{0 +}  = n_{0}$. Then the parameters in the
doped and undoped regions are related by the integrals of Eqs.
(\ref{eq9})-(\ref{eq11}) according to
\begin{equation}
\label{eq13}
 - U_{p} + \mu _{h1} E_{1} = 0,
\end{equation}
\begin{equation}
\label{eq14}
 - n_{1 +}  U_{p} - n_{1 +}  \mu _{ +}  E_{1} = - n_{0} U_{p} - n_{0} \mu _{
+}  E_{0} ,
\end{equation}
\begin{equation}
\label{eq15}
 - n_{1 -}  U_{p} + n_{1 -}  \mu _{ -}  E_{1} = - n_{0} U_{p} + n_{0} \mu _{
-}  E_{0} ,
\end{equation}
Equation (\ref{eq13}) specifies the electric field in the doped region as
$E_{1} = U_{p} / \mu _{h1} $, which may be neglected taking into
account high mobility of the holes in comparison to the ions in the doped
material: $E_{1} = U_{p} / \mu _{h1} \propto (\mu _{\pm}  / \mu
_{h1} )E_{0} \ll E_{0} $. Equation (\ref{eq12}) in the uniform doped
region determines zero net charge as $n_{1 -}  - n_{1 +} = n_{1h} $.
Then equations (\ref{eq14}), (\ref{eq15}) specify the
front velocity and ion concentration in the doped region as
\begin{equation}
\label{eq16}
U_{p} = {\frac{{n_{0}}} {{n_{1h}}} }(\mu _{ +}  + \mu _{ -}  )E_{0} ,
\end{equation}
\begin{equation}
\label{eq17}
n_{1 -}  = n_{0} + n_{1h} {\frac{{\mu _{ -}} } {{\mu _{ +}  + \mu _{ -}
}}},
\end{equation}
\begin{equation}
\label{eq18}
n_{1 +}  = n_{0} - n_{1h} {\frac{{\mu _{ +}} } {{\mu _{ +}  + \mu _{ -}
}}}.
\end{equation}
Similar formulas may be obtained for the n-doping front, i.e.
\begin{equation}
\label{eq19}
U_{n} = {\frac{{n_{0}}} {{n_{2e}}} }(\mu _{ +}  + \mu _{ -}  )E_{0} ,
\end{equation}
\begin{equation}
\label{eq20}
n_{2 +}  = n_{0} + n_{2e} {\frac{{\mu _{ +}} } {{\mu _{ +}  + \mu _{ -}
}}},
\end{equation}
\begin{equation}
\label{eq21}
n_{2 -}  = n_{0} - n_{2e} {\frac{{\mu _{ -}} } {{\mu _{ +}  + \mu _{ -}
}}},
\end{equation}
where the label ``2'' designates the conducting substance behind the n-doping front.
We note that the p- and n-type doping fronts propagate with different velocities related by
\begin{equation}
\label{eq22}
U_{n} = {\frac{{n_{1h}}} {{n_{1e}}} }U_{p} = \beta U_{p} .
\end{equation}
Taking data for electron and hole concentration $n_{1h} = 8.6 \cdot
10^{25}m^{ - 3}$, $n_{1e} = 1.3 \cdot 10^{26}m^{ - 3}$ obtained in
the experiments  \cite{Matyba-09,Robinson,Modestov-10} we find
the velocity of the n-front to be less than that of the p-front with $\beta =
0.661$.

Next we consider a semiconductor of finite size with a voltage $\phi
_{0} $ applied to the end electrodes and two planar doping fronts moving
towards each other. The initial distance between the fronts is $L_{0} $.
Since the distance between the fronts $L(t)$ decreases in time and
the potential difference is fixed, then electric field $E_{0} (t)$
in the gap between the fronts grows in time and the fronts
accelerate according to Eqs. (\ref{eq16}), (\ref{eq19}). We
designate positions of the fronts by $X_{p} (t)$, $X_{n} (t)$. In
experiments \cite{Matyba-09,Robinson,Modestov-10}, the n-front
(electron doping) starts later, after a time delay $t_{e} $;  we thus have
two time intervals in the solution: $t < t_{e} $ and $t > t_{e} $.
We start with the first interval, $t < t_{e} $, when only the
p-front propagates. In that case Eq. (\ref{eq16}) for the p-front
velocity is reduced to
\begin{equation}
\label{eq23}
{\frac{{dX_{p}}} {{dt}}} = {\frac{{n_{0}}} {{n_{1h}}} }(\mu _{ +}  + \mu _{
-}  ){\frac{{\phi _{0}}} {{L}}} = {\frac{{U_{p0}}} {{1 - X_{p} / L_{0}}} }
\end{equation}
where $U_{p0} $ is the initial velocity of the doping
front. Integrating (\ref{eq23}) we find
\begin{equation}
\label{eq24}
{\frac{{X_{p}}} {{L_{0}}} } = 1 - \sqrt {1 - 2U_{p0} t / L_{0}}  .
\end{equation}
At the moment $t_{e} $, when the n-front starts, the p-front is already at
the position $X_{p} (t_{e} ) / L_{0} = 1 - \sqrt {1 - 2U_{p0} t_{e} / L_{0}
} $, and we have  the reduced distance between the fronts $L_{e} = L(t_{e} ) = L_{0}
- X_{p} (t_{e} )$. Next, we consider the interval $t > t_{e} $, when both
fronts propagate. In that case the distance between the fronts varies as $L
= X_{n} - X_{p} $ with
\begin{equation}
\label{eq25}
{\frac{{dX_{p}}} {{dt}}} = {\frac{{U_{p0}}} {{L / L_{0}}} },
\quad
{\frac{{dX_{n}}} {{dt}}} = - {\frac{{\beta U_{p0}}} {{L / L_{0}}} },
\end{equation}
so that
\begin{equation}
\label{eq26}
{\frac{{dL}}{{dt}}} = - (1 + \beta ){\frac{{U_{p0}}} {{L / L_{0}}} }.
\end{equation}
Then, for $t > t_{e} $, we obtain the solution
\begin{equation}
\label{eq27} L(t) = \sqrt {L_{e} ^{2} - 2(1 + \beta )U_{p0} L_{0} (t
- t_{e} )} ,
\end{equation}
\begin{equation}
\label{eq28} X_{p} = X_{p} \left( {t_{e}}  \right) + {\frac{1}{1 +
\beta} }{\left[ {L_{e} - L(t)} \right]},
\end{equation}
\begin{equation}
\label{eq29} X_{n} = L_{0} - {\frac{{\beta}} {{1 + \beta}} }{\left[
{L_{e} - L(t)} \right]}.
\end{equation}
%\begin{equation}
%\label{eq28} X_{p} = X_{p} \left( {t_{e}}  \right) + {\frac{1}{1 +
%\beta} }{\left[ {L_{e} - \sqrt {L_{e} ^{2} - 2(1 + \beta )U_{p0}
%L_{0} (t - t_{e} )}} \right]},
%\end{equation}
%\begin{equation}
%\label{eq29}
%X_{n} = L_{0} - {\frac{{\beta}} {{1 + \beta}} }{\left[ {L_{e} - \sqrt {L_{e}
%^{2} - 2(1 + \beta )U_{p0} L_{0} (t - t_{e} )}}  \right]}.
%\end{equation}
\begin{figure}
\includegraphics[width=3.6in,height=2.8in]{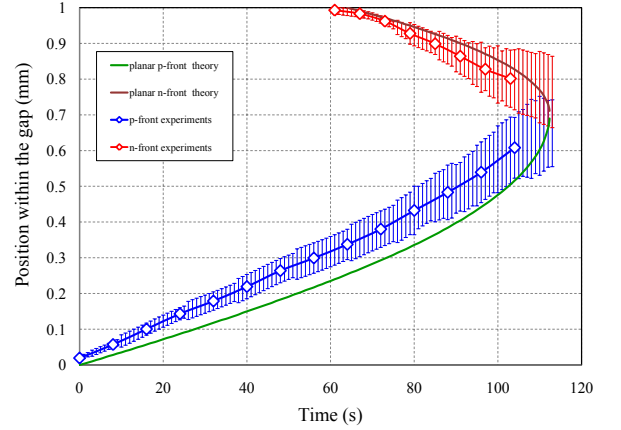}
\caption{Positions of the doping fronts versus time as predicted by the analytical theory (solid lines) and obtained experimentally (markers).
The uncertainty bars indicate the difference between the fastest and the slowest parts in the multidimensional front structure.}
\end{figure}
Figure 2 shows a comparison of the positions of the planar p-
and n-fronts, as found analytically through Eqs. (\ref{eq28}), (\ref{eq29}), to the experimental
data. The markers show the average position of the fronts; the error
bars indicate the difference between the fastest and the slowest parts of the two-dimensional (2D) front structure (width of
the 2D  front brush). Such 2D structures develop due to front instabilities,
which are beyond the scope of the present paper and which will be
presented elsewhere \cite{Bychkov-sub}. Figure 2 indicates that the
analytical 1D solution describes the dynamics of the backside of the front
structure quite well (within the 10\%
accuracy of the experimental data). At the same time, the leading parts of the front move
noticeably faster than the 1D theory predicts. The last fact
demonstrates an important role of the 2D instabilities in the doping
front dynamics, which is the subject for the future
work.

\section{Internal structure of a doping front}

\subsection{Basic equations}

In the present section we investigate the internal structure of a stationary
doping front driven by a constant uniform external electric field $E_{0} $.
We start by studying  the p-front, since analysis of the n-front is
similar. The internal structure of the doping front is controlled by diffusion
in Eqs. (\ref{eq9})-(\ref{eq11}) integrated within the front
\begin{equation}
\label{eq30}
 - n_{h} U_{p} + n_{h} \mu _{h} (E - E_{N} ) = D_{h} {\frac{{dn_{h}
}}{{dx}}},
\end{equation}
\begin{equation}
\label{eq31}
(n_{0} - n_{ -}  )U_{p} - \mu _{ -}  (n_{ -}  E - n_{0} E_{0} ) = D_{ -}
{\frac{{dn_{ -}} } {{dx}}},
\end{equation}
\begin{equation}
\label{eq32}
(n_{0} - n_{ +}  )U_{p} + \mu _{ +}  (n_{ +}  E - n_{0} E_{0} ) = D_{ +}
{\frac{{dn_{ +}} } {{dx}}}.
\end{equation}
A characteristic length scale $L_{p} $ involved in the problem is determined
by the slow diffusion of ions, thus we may choose, e.g., $D_{ -}  $ to define
\begin{equation}
\label{eq33}
L_{p} \equiv {\frac{{D_{ -}} } {{U_{p}}} } = {\frac{{D_{ -}  n_{1h}}} {{(\mu _{ +
} + \mu _{ -}  )n_{0} E_{0}}} }.
\end{equation}
As an example, taking the experimental data of Refs. \cite{Matyba-09,Robinson,Modestov-10} for
$n_{1h} = 8.2 \cdot 10^{25}{\rm m}^{ - {\rm 3}}$, $n_{0} = 3.1 \cdot
10^{26}{\rm m}^{ - {\rm 3}}$, $\mu _{ +}  = 1.0 \cdot 10^{ - 10}{\rm
m}^{{\rm 2}}{\rm /} {\rm V}{\rm s}$, $\mu _{ -}  = 2.2 \cdot 10^{ - 10}{\rm
m}^{{\rm 2}}{\rm /} {\rm V}{\rm s}$, $T = 360{\rm K}$, $E_{0} = 3 \cdot
10^{3}{\rm V}{\rm /} {\rm m}$, we can evaluate the characteristic initial p-front
velocity as $U_{p0} \approx 3.6 \cdot 10^{ - 6}{\rm m}{\rm /} {\rm s}$ and
the initial length scale as $L_{p0} \approx 1.8 \cdot 10^{ - 6}{\rm m}$.
Still, one has to remember that $L_{p} $ does not portray the full structure
of the doping front. We show below that there are several different
characteristic length scales within the front from the undoped to the doped
region.

We may also estimate the typical deviations from quasi-neutrality within the
front through
\begin{equation}
\label{eq34}
{\frac{{n_{ +}  - n_{ -}  + n_{h}}} {{n_{0}}} } \propto {\frac{{\varepsilon
_{0}}} {{en_{0}}} }{\frac{{dE}}{{dx}}} \propto {\frac{{\varepsilon _{0}
E_{0}}} {{eL_{p0} n_{0}}} } \approx 2.9 \cdot 10^{ - 10} \ll 1,
\end{equation}
which allows replacing the Poisson equation Eq. (\ref{eq12}) by the condition of
quasi-neutrality
\begin{equation}
\label{eq35}
n_{ -}  - n_{ +}  = n_{h}
\end{equation}
everywhere within the front with very good accuracy. As a
comparison, the electrochemical fronts in the ambipolar devices
involve noticeable deviations from quasi-neutrality as shown in Ref.
\cite{Wang-09}.

The next important question concerns the terms specifying the injection barrier,
$\phi _{N} $, $E_{N} $, in Eqs. (\ref{eq5}) and (\ref{eq30}). Below, we demonstrate  that it is
indeed impossible to describe the doping front structure without such a barrier term.
Let us consider the doped region at the back side of the front with
$\mu _{h} \gg \mu _{\pm}  $. Without the barrier term, one has the
force-balance equation
\begin{equation}
\label{eq36}
n_{h} m{\frac{{d{\rm {\bf v}}}}{{dt}}} = - en_{h} \nabla \phi - k_{B}
T\nabla n_{h} - {\frac{{n_{h} m}}{{\tau}} }{\rm {\bf v}}.
\end{equation}
for holes similar to Eq. (\ref{eq1}).
Within the drift-diffusion approximation, the inertial term $n_{h} md{\rm
{\bf v}} / dt$ in Eq. (\ref{eq36}) is negligible. The characteristic velocities of the
doping process are controlled by relatively low ion mobility, Eq. (\ref{eq16}).
Therefore, by order-of-magnitude the collision term in Eq. (\ref{eq36}) is also
negligible in comparison with the electric term in the doped region
\begin{eqnarray}
\label{eq37}
&&\!\!\!\!\!\! {\frac{{n_{h} m}}{{\tau}} }v \propto {\frac{{m}}{{\tau}} }n_{0} (\mu _{ +}
+ \mu _{ -}  )\nabla \phi \propto
\nonumber \\ &&
{\frac{{\mu _{ +}  + \mu _{ -}} } {{\mu
_{h}}} }en_{h} \nabla \phi \ll en_{h} \nabla \phi .
\end{eqnarray}
Then, in the doped region, Eq. (\ref{eq36}) should describe
hydrostatic equilibrium for the holes, with balance between electric and pressure forces
proportional to $ - \nabla \phi $ and $ - \nabla n_{h} $,
respectively. However, in the geometry of the doping process, the
electric and pressure forces point in the same direction (e.g. to
the right in Fig. 1) and therefore \textit{cannot} balance each other.
Without the barrier term, Eq. (\ref{eq36}) suggests that holes are freely accelerated into the semiconductor, which contradicts the
very essence of the doping process. In order to obtain a doping
front, one has to balance the electric and pressure forces in Eq.
(\ref{eq36}) by a counter-term, which takes into account the
thermodynamic barrier of the doping process
\begin{equation}
\label{eq38}
n_{h} m{\frac{{d{\rm {\bf v}}}}{{dt}}} = - en_{h} \nabla (\phi - \phi _{N} )
- k_{B} T\nabla n_{h} - {\frac{{n_{h} m}}{{\tau}} }{\rm {\bf v}}.
\end{equation}
It has been suggested in Ref. \cite{Modestov-10} to consider $\phi _{N} $ in the form of
the Nernst potential, given by
\begin{equation}
\label{eq39}
\phi _{N} (n_{h} ,T) = {\frac{{k_{B} T}}{{e}}}\ln \left( {{\frac{{n_{h}
}}{{n_{h,\infty}  - n_{h}}} }} \right),
\end{equation}
where $n_{h,\infty}  $ is the maximal possible concentration of
holes. In general, $n_{h\infty}  $ may be larger than the value
$n_{h1} $ of the experimentally observed hole concentration  behind the p-doping
front. For this reason, in the present paper we introduce also a numerical
parameter $f$ as $n_{h\infty}  \equiv fn_{h1} $ with $f > 1$. We will show that qualitative properties of the doping front do not depend on the parameter $f$.
As
mentioned above, the Nernst potential should be viewed as an
injection barrier, and is only described as an electrostatic
contribution out of convention. Moreover, the important dynamical
aspects of the Nernst potential enter Eq. (\ref{eq36}) in the form
of a gradient. Since the inclusion of the Nernst potential is
formally valid only in the highly doped region, it can therefore be
set to an arbitrary constant value in the undoped region, i.e.
\begin{equation}
\label{eq40}
\phi _{N} (n_{h} ,T) = 0.
\end{equation}
A continuous description of the doping front structure also requires
a continuous change of the Nernst potential from Eq. (\ref{eq39}) to
Eq. (\ref{eq40}). Since, unfortunately, there is no theoretical
thermodynamic model describing such a transition at present, we
 introduce a dimensionless phenomenological function $\psi $ with the
injection barrier potential given by
\begin{equation}
\label{eq41}
\phi _{N} (n_{h} ,T) = \psi {\frac{{k_{B} T}}{{e}}}\ln \left( {{\frac{{n_{h}
}}{{n_{h,\infty}  - n_{h}}} }} \right),
\end{equation}
where $\psi = 1$ in the doped region and $\psi = 0$ in the undoped region. A
more detailed form of the function $\psi $ will be discussed below. Taking
into account the Nernst potential, Eq. (\ref{eq30}) is modified in the doped region
according to
\begin{equation}
\label{eq42}
 - n_{h} U_{p} + n_{h} \mu _{h} E = D_{h} \left( {1 - \psi {\frac{{fn_{h1}
}}{{fn_{h1} - n_{h}}} }} \right){\frac{{dn_{h}}} {{dx}}}.
\end{equation}
The set of Eqs. (\ref{eq31}), (\ref{eq32}), (\ref{eq35}),
(\ref{eq42}) determine the internal structure of the p-doping front.

Another important feature of the system is a strongly nonlinear
dependence of the hole mobility on concentration $\mu _{h} (n_{h} )$
found experimentally \cite{Shimotani}.  Because of this dependence, hole mobility is
much larger than the ion mobility in the doped region $\mu _{h} \gg\mu _{\pm} $, but it becomes much lower $\mu _{h} \ll \mu _{\pm}
$ in the undoped region where $n_{h} / n_{h1} \ll 1$. The nonlinear
dependence is consistent with basic understanding of doping as the
process of increasing mobility of charge carriers. Reference
\cite{Modestov-10} makes use of the following empirical fit for the hole mobility
\begin{equation}
\label{eq43}
\mu _{h} = 3.85\times 10^{ - 8}{\left[ {1 + \tanh \left( {26.6{\frac{{n_{h}
}}{{n_{0}}} } - 4.3} \right)} \right]},\, {\rm m}^{{\rm 2}}{\rm /} {\rm
V}\,{\rm s}.
\end{equation}
A similar property holds
also for electrons.

\subsection{Dimensionless equations}

In order to simplify the analysis, we introduce dimensionless variables for the coordinate, concentrations and
electric field: $\xi = x / L_{p} $, $\alpha _{\pm} = n_{\pm} / n_{0} $, $\alpha _{h} = n_{h} / n_{0} $,
$\;\varepsilon = E / E_{0} $. Furthermore, we  introduce two parameters specifying ratio of ion mobilities and the
front velocity according to
\begin{equation}
\label{eq44}
\delta = \mu _{ -}/\mu _{ +},
\quad
C = \mu _{ -}  E_{0}/ U_{p0}.
\end{equation}
In particular, we have $\delta = 2.2$ for the active material used in
the experiments of Refs. \cite{Matyba-09,Robinson,Modestov-10}. In
the dimensionless form, the governing equations (\ref{eq42}),
(\ref{eq31}), (\ref{eq32}), (\ref{eq35}) become
\begin{equation}
\label{eq45}
\gamma _{eff} {\frac{{d\alpha _{h}}} {{d\xi}} }= \alpha _{h} \left( {CM_{h}
\varepsilon - 1} \right),
\end{equation}
\begin{equation}
\label{eq46}
{\frac{{d\alpha _{ -}} } {{d\xi}} } = 1 + C - \alpha _{ -}  \left( {1 +
C\varepsilon}  \right),
\end{equation}
\begin{equation}
\label{eq47}
{\frac{{d\alpha _{ +}} } {{d\xi}} } = \delta - C - \alpha _{ +}  \left(
{\delta - C\varepsilon}  \right),
\end{equation}
\begin{equation}
\label{eq48}
\alpha _{ +}  - \alpha _{ -}  + \alpha _{h} = 0,
\end{equation}
where the dimensionless hole mobility and the effective
dimensionless hole diffusion are
\begin{equation}
\label{eq49}
M_{h} = 175{\left[ {1 + \tanh \left( {26.6\alpha _{h} - 4.3} \right)}
\right]},
\end{equation}
\begin{equation}
\label{eq50}
\gamma _{eff} = M_{h} \left( {1 - {\frac{{\psi \alpha _{h\infty}} } {{\alpha
_{h\infty}  - \alpha _{h}}} }} \right).
\end{equation}

The numerical solution also requires  boundary conditions in the doped region,
as specified in Sec. III under a simplifying (though realistic)
assumption of infinitely large hole mobility in the doped region, $\mu _{h1}
\gg \mu _{\pm}  $. Still, it is also useful to determine the boundary
conditions without employing such an assumption. The hole concentration in
the doped region is known from experiments. Thus, we have to find the ion
concentrations in the doped region and the exact value of the $C$-parameter in Eq. (\ref{eq44}),
taking into account a finite (though small) electric field $\varepsilon _{1}
$. Since all derivatives in Eqs. (\ref{eq45})-(\ref{eq47}) are zero in the doped region, then
we obtain the following boundary conditions
\begin{equation}
\label{eq51}
\varepsilon _{1} = {\frac{{1}}{{CM_{1h}}} },
\quad
\alpha _{1 -}  = {\frac{{1 + C}}{{1 + C\varepsilon _{1}}} },
\quad
\alpha _{1 +}  = {\frac{{\delta - C}}{{\delta - C\varepsilon _{1}}} }{\rm
.}
\end{equation}
Substituting (\ref{eq51}) into (\ref{eq48}) we find the parameter
$C$
\begin{equation}
\label{eq52}
C = {\frac{{\alpha _{1h}}} {{\left( {\delta + 1} \right)}}}\left( {\delta -
{\frac{{1}}{{M_{1h}}} }} \right)\left( {{\frac{{1}}{{M_{1h}}} } + 1} \right)
+ {\frac{{1}}{{M_{1h}}} },
\end{equation}
where $\alpha _{1h} = n_{1h} / n_{0} = 0.277$ determined experimentally. In the limit of infinitely high mobility of holes in the doped
region, $M_{1h} \gg 1$, Eq. (\ref{eq52}) goes over to Eq. (\ref{eq16}), which may be written
in the dimensionless form as $C = \alpha _{1h} \delta / (\delta + 1)$. For the
experimental values of the hole mobility, the difference between Eqs. (\ref{eq16})
and (\ref{eq52}) is about 2\% . Boundary conditions in the undoped region are, by
definition, $\varepsilon _{0} = 1$, $\alpha _{0 -}  = \alpha _{0 +}  = 1$,
$\alpha _{0h} = 0$.

\subsection{Asymptotic behavior in the doped region}

We next consider the asymptotic behavior in the specific zones of the front: in the doped region, the undoped region and the transition point between doped and undoped matter. We
 start with the doped region. Within the limit of high hole mobility, $\mu _{h1} \gg \mu _{\pm}  $ i.e. $M_{h1} \gg 1$, we have $\varepsilon _{1} = 0$, $\alpha
_{h} = \alpha _{1h} $, $\alpha _{1 -}  = 1 + \alpha _{1h} \delta / (\delta +
1)$, $\alpha _{1 +}  = 1 - \alpha _{1h} / (\delta + 1)$ in the doped region
at $\xi = - \infty $. We are interested in the asymptotic approach to the
doped state, and we investigate small deviations from the limiting values
$\varepsilon _{1} = \tilde {\varepsilon} $, $\alpha _{h} = \alpha _{1h} +
\tilde {\alpha} _{h} $, $\alpha _{ -}  = \alpha _{1 -}  + \tilde {\alpha} _{
-}  $, $\alpha _{ +}  = \alpha _{1 +}  + \tilde {\alpha} _{ +}  $. We also
take into account the Nernst term ($\psi = 1)$ in the doped region, which
leads to
\begin{equation}
\label{eq53}
\gamma _{eff} = - M_{h} {\frac{{\alpha _{h}}} {{f\alpha _{1h} - \alpha _{h}
}}}
\end{equation}
in the definition Eq. (\ref{eq50}). We immediately see
that, in the case of $f = 1$, $\gamma _{eff} $ diverges in the doped region as $\xi \to -
\infty $. Thus, we have to consider two
separate cases of $f > 1$ and $f = 1$ yielding an exponential and
a power-law approach to the limiting values, respectively.

We begin with the first case, $f > 1$, for which $\gamma _{eff} $ is a
constant coefficient in Eqs.
(\ref{eq45})-(\ref{eq48}), $\gamma _{eff} = - M_{h} / (f - 1)$. Linearizing  Eqs. (\ref{eq45})-(\ref{eq48})  with respect to
small deviations in the doped region, we obtain
\begin{equation}
\label{eq54}
 - {\frac{{1}}{{f - 1}}}{\frac{{d\tilde {\alpha} _{h}}} {{d\xi}} } =
{\frac{{ \alpha _{1h}^{2}\delta}} {{\delta + 1}}}\tilde {\varepsilon} {\rm
,}
\end{equation}
\begin{equation}
\label{eq55}
{\frac{{d\tilde {\alpha} _{ -}} } {{d\xi}} } = - \tilde {\alpha} _{ -}  -
\left( {1 + {\frac{{ \alpha _{1h}\delta}} {{\delta + 1}}}}
\right){\frac{{ \alpha _{1h}\delta}} {{\delta + 1}}}\tilde {\varepsilon
},
\end{equation}
\begin{equation}
\label{eq56}
{\frac{{d\tilde {\alpha} _{ +}} } {{d\xi}} } = -  \tilde {\alpha} _{ +
}\delta + \left( {1 - {\frac{{\alpha _{1h}}} {{\delta + 1}}}}
\right){\frac{{\alpha _{1h} \delta}} {{\delta + 1}}}\tilde {\varepsilon
},
\end{equation}
\begin{equation}
\label{eq57}
\tilde {\alpha} _{ +}  - \tilde {\alpha} _{ -}  + \tilde {\alpha} _{h} =
0.
\end{equation}
The system of Eqs. (\ref{eq54})-(\ref{eq57}) has an exponential
solution in the form $\tilde {\varepsilon}  \propto \tilde {\alpha}
_{h} \propto \tilde {\alpha} _{ -} \propto \tilde {\alpha} _{ +}
\propto \exp (\chi \xi )$ with positive eigenvalues $\chi $
corresponding to the exponent decay at $\xi \to - \infty $. In
general, the eigenvalue may be calculated numerically. Still, we can
find an analytical solution to the system in the most important
limit of $f - 1 \ll 1$, which describes the most interesting
features of the front in the doped region. Substituting
the deviations of electric field $\tilde {\varepsilon} $ from Eq.
(\ref{eq54}) into Eqs. (\ref{eq55}) and (\ref{eq56}), we obtain
\begin{equation}
\label{eq58}
{\frac{{d\tilde {\alpha} _{ -}} } {{d\xi}} } = - \tilde {\alpha} _{ -}  +
\left( {1 + {\frac{{ \alpha _{1h}\delta}} {{\delta + 1}}}}
\right){\frac{{1}}{{(f - 1)\alpha _{1h}}} }{\frac{{d\tilde {\alpha} _{h}
}}{{d\xi}} },
\end{equation}
\begin{equation}
\label{eq59}
{\frac{{1}}{{\delta}} }{\frac{{d\tilde {\alpha} _{ +}} } {{d\xi}} } = -
\tilde {\alpha} _{ +}  - {\frac{{1}}{{\delta}} }\left( {1 - {\frac{{\alpha
_{1h}}} {{\delta + 1}}}} \right){\frac{{1}}{{(f - 1)\alpha _{1h}
}}}{\frac{{d\tilde {\alpha} _{h}}} {{d\xi}} }.
\end{equation}
Thus, in the case of $f - 1 \ll 1$, taking the difference of Eqs. (\ref{eq55}) and
(\ref{eq56}) and accounting for Eq. (\ref{eq57}) we find that
\begin{equation}
\label{eq60}
\tilde {\alpha} _{h} = {\frac{{\delta + 1 + \alpha _{1h} (\delta -
1)}}{{\delta (f - 1)\alpha _{1h}}} }{\frac{{d\tilde {\alpha} _{h}}} {{d\xi
}}},
\end{equation}
with the eigenvalue
\begin{equation}
\label{eq61}
\chi = {\frac{{\delta \alpha _{1h} (f - 1)}}{{\delta + 1 + \alpha _{1h}
(\delta - 1)}}}.
\end{equation}
For the experimental data $\alpha _{1h} = 0.277$ and $\delta = 2.2$ we find
that relaxation of the parameters to the saturation values in the
p-doped region occurs on a length scale $ \approx 7.1L_{p} / (f - 1)$.
Therefore, in the limit of $f - 1 \ll 1$, the relaxation happens on
 length scales much greater than the characteristic length $L_{p} $, Eq.
(\ref{eq33}), related to ion diffusion. We stress also that without employing the
injection barrier in the doped region  [i.e. taking $\psi = 0$ in Eq. (\ref{eq40})],
we do not find any deviation mode vanishing asymptotically to $\xi \to -
\infty $ in agreement with the previous numerical results \cite{Modestov-10}. The physical
meaning of this effect was explained above in Sec. IV A.

The relaxation process goes even slower in the specific case of $f = 1$,
when Eq. (\ref{eq54}) becomes   intrinsically nonlinear
\begin{equation}
\label{eq62}
{\frac{{1}}{{\tilde {\alpha} _{h}}} }{\frac{{d\tilde {\alpha} _{h}}} {{d\xi
}}} = {\frac{{ \alpha _{1h}\delta}} {{\delta + 1}}}\tilde {\varepsilon} {\rm
.}
\end{equation}
Then, instead of Eq. (\ref{eq60}), we obtain
\begin{equation}
\label{eq63}
{\frac{{d\tilde {\alpha} _{h}}} {{d\xi}} } = - {\frac{{ \tilde {\alpha
}_{h}^{2}\delta}} {{\delta + 1 + \alpha _{1h} (\delta - 1)}}},
\end{equation}
with the asymptotic solution
\begin{equation}
\label{eq64}
\tilde {\alpha} _{h} = {\frac{{\delta + 1 + \alpha _{1h} (\delta -
1)}}{{ \xi \delta}} }
\end{equation}
for $\xi \to - \infty $. According to Eq. (\ref{eq64}), in the case of $f = 1$ the relaxation of the hole
concentration goes inversely proportional to the distance from the doping
front. Such behavior is much slower than the exponential law predicted by
Eq. (\ref{eq60}) for $f > 1$.

Thus, we obtain a smooth relaxation of the p-front parameters to the final
values in the doped region on the length scales much larger than $L_{p} $,
Eq. (\ref{eq33}), determined by ion diffusion.

\subsection{Asymptotic behavior in the undoped region}

In this subsection we study the asymptotic front structure in the undoped
region ($\xi \to \infty )$ with $\varepsilon _{0} = 1$, $\alpha _{0 -}  =
\alpha _{0 +}  = 1$, $\alpha _{0h} = 0$, $\psi = 0$ in the limit $\gamma
_{eff} = M_{h0} \ll 1$, the latter due to the strongly nonlinear dependence of hole
mobility on concentration, Eq. (\ref{eq43}). Using again a tilde to denote the deviation variables,  Eqs. (\ref{eq45})-(\ref{eq48}) may be linearized with
respect to small deviations to give
\begin{equation}
\label{eq65}
M_{h0} {\frac{{d\tilde {\alpha} _{h}}} {{d\xi}} } = - \tilde {\alpha} _{h}
,
\end{equation}
\begin{equation}
\label{eq66}
{\frac{{d\tilde {\alpha} _{ -}} } {{d\xi}} } = - \tilde {\alpha} _{ -}
\left( {1 + {\frac{{\alpha _{1h} \delta}} {{\delta + 1}}}} \right) -
{\frac{{\alpha _{1h} \delta}} {{\delta + 1}}}\tilde {\varepsilon} ,
\end{equation}
\begin{equation}
\label{eq67}
{\frac{{d\tilde {\alpha} _{ +}} } {{d\xi}} } = - \tilde {\alpha} _{ +}
\left( {\delta - {\frac{{\alpha _{1h} \delta}} {{\delta + 1}}}} \right) +
{\frac{{\alpha _{1h} \delta}} {{\delta + 1}}}\tilde {\varepsilon} ,
\end{equation}
\begin{equation}
\label{eq68}
\tilde {\alpha} _{ +}  - \tilde {\alpha} _{ -}  + \tilde {\alpha} _{h} =
0.
\end{equation}
The system (\ref{eq65})-(\ref{eq68}) has two independent modes in
the form of $\tilde {\varepsilon}  \propto \tilde {\alpha} _{h}
\propto \tilde {\alpha} _{ -} \propto \tilde {\alpha} _{ +}  \propto
\exp (\chi \xi )$ decaying exponentially at $\xi \to \infty $
with $\chi < 0$. One mode is related to the perturbations of the holes,
$\tilde {\alpha} _{h} \ne 0$, with the eigenvalue
\begin{equation}
\label{eq69}
\chi = - 1 / M_{h0} ,
\end{equation}
 which is obtained in a straightforward way from Eq. (\ref{eq65}).
In the limit of ultimately low hole mobility in the undoped region,
$M_{h0} \ll 1$, this mode is characterized by extremely sharp
gradients, thus leading to a steep head of the front with an associated
length scale $ \approx M_{h0} L_{p} $, which is much smaller than the length
scale $L_{p} $ due to ion diffusion. This mode is also expected to lead
to a sharp peak in the electric field, which is proportional to the
gradient of hole concentration according to
\begin{equation}
\label{eq70}
{\frac{{d\tilde {\alpha} _{h}}} {{d\xi}} } = - 2{\frac{{\alpha _{1h} \delta
}}{{\delta + 1}}}\tilde {\varepsilon} .
\end{equation}

The second mode in the undoped region happens with  zero deviations of
the hole concentration, $\tilde {\alpha} _{h} = 0$, and equal non-zero
deviations for the ion concentrations, $\tilde {\alpha} _{ -}  = \tilde
{\alpha} _{ +}  \ne 0$. Then Eqs. (\ref{eq66}), (\ref{eq67}) yield
\begin{equation}
\label{eq71}
{\frac{{d\tilde {\alpha} _{\pm}} } {{d\xi}} } = - {\frac{{1 + \delta
}}{{2}}}\tilde {\alpha} _{\pm}
\end{equation}
with the eigenvalue
\begin{equation}
\label{eq72}
\chi = - {\frac{{1 + \delta}} {{2}}}.
\end{equation}
The second mode is controlled by ion diffusion with the typical
length scale comparable to $L_{p} $. Because of the second mode one
should expect a non-monotonic behavior of the ion concentrations in the
doped region. After sharp changes related to the first mode with
complete vanishing of the hole concentration, a relatively slow relaxation
of ions to $\alpha _{0 -}  = \alpha _{0 +}  = 1$ is thus expected.

\subsection{Behavior in the transition point}

The solution to the system (\ref{eq45})-(\ref{eq48}) demonstrates that the doping front
possesses one more specific point, which indicates transition from the doped
to undoped zones. This point gives the answer to the question
where the injection barrier (the Nernst term) should be switched off via
the phenomenological parameter $\psi $ going over from 1 to 0. \textit{A priori}, it is
natural to expect the critical point to be in the region where the hole mobility
becomes comparable to the ion mobility. However, it turns out that the
system specifies an \textit{exact} position of the critical transition point (which we label by $c$)
characterized by zero derivatives of Eqs. (\ref{eq45})-(\ref{eq48}), i.e.
\begin{equation}
\label{eq73}
0 = {\frac{{ \alpha _{1h}\delta}} {{\delta + 1}}}M_{h} (\alpha _{hc}
)\varepsilon _{c} - 1,
\end{equation}
\begin{equation}
\label{eq74}
0 = 1 + {\frac{{ \alpha _{1h}\delta}} {{\delta + 1}}} - \alpha _{c -}
\left( {1 + {\frac{{ \alpha _{1h}\delta}} {{\delta + 1}}}\varepsilon _{c}}
\right),
\end{equation}
\begin{equation}
\label{eq75}
0 = \delta - {\frac{{ \alpha _{1h}\delta}} {{\delta + 1}}} - \alpha _{c +}
\left( {\delta - {\frac{{ \alpha _{1h}\delta}} {{\delta + 1}}}\varepsilon
_{c}}  \right),
\end{equation}
\begin{equation}
\label{eq76}
\alpha _{c +}  - \alpha _{c -}  + \alpha _{hc} = 0.
\end{equation}
Due to the essentially nonlinear dependence of hole mobility on
concentration, see Eq. (\ref{eq43}), the set of equations
(\ref{eq73})-(\ref{eq76}) is also strongly nonlinear and it can only be
solved  numerically. Thus, solving Eqs. (\ref{eq73})-(\ref{eq76}) for
the experimentally obtained parameters $\alpha _{1h} = 0.277$ and
$\delta = 2.2$, we  find the critical point at $\alpha _{hc}
\approx 0.092$, $M_{h} (\alpha _{hc} ) = \mu _{h} / \mu _{ -}
\approx 8.5$ with the scaled electric field at that point calculated  as $\varepsilon _{c} \approx 0.65$ from
Eq. (\ref{eq73}).

The critical point has an interesting physical meaning. Going back to the
dimensional equations for the front structure, Eqs. (\ref{eq30})-(\ref{eq32}), and setting
all derivatives equal to zero, we find the critical point corresponding to
\begin{equation}
\label{eq77}
 - U_{p} + \mu _{h} E = 0.
\end{equation}
Thus, at the critical point we obtain holes moving locally with the
same velocity $\mu _{h} E$ as the p-doping front, which may be also
interpreted as a ``resonance'' between the light charges and the
front. To the left of this point (in the doped region) hole mobility
may provide faster velocity of the particles in comparison with the
front. To the right of the critical point, hole mobility is too low
for holes to keep up with the front velocity. Therefore the critical point
plays the natural role of a border between the doped and undoped
regions.

\subsection{Numerical solution for the front structure}

\begin{figure}
\includegraphics[width=3.6in,height=2.8in]{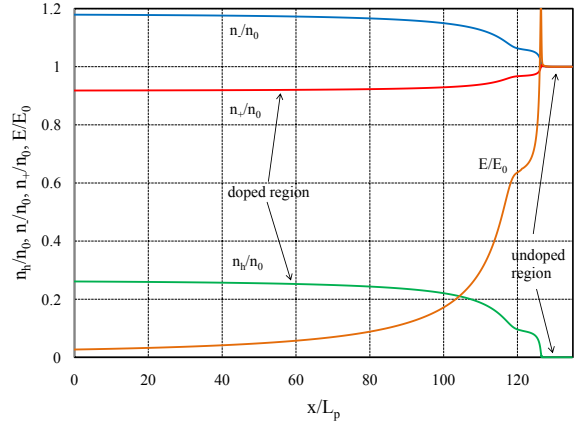}
\caption{Concentrations of holes and ions, and the electric field inside the p-doping front; $f = 1.1.$}
\end{figure}

In this section, we solve Eqs. (\ref{eq45})-(\ref{eq50}) numerically with the boundary conditions ahead of
the doping front and behind the front as obtained in Sec. IV B. Figure 3
shows the internal structure of a stationary p-doping front in terms of  the
normalized concentrations of holes, positive and negative ions and the electric field as a function of distance.
The front propagates to the right converting the undoped semiconductor
material with low conductivity to the doped one with high conductivity.
In agreement with the
analysis of Subsections IV C-E, the whole doping front has a complicated
nonlinear structure with several specific zones characterized by different
length scales. The length scale variations within the front are related, first
of all, to dramatic changes of the hole mobility by three orders of magnitude
from the undoped to doped regions. We point out a sharp head of the
doping front in the undoped region, an extremely smooth relaxation tail
in the doped region and an additional plateau connecting these two
regions. The length scale variations inside the doping front resemble a similar effect encountered in laser deflagration, where the length scale may also change  by several orders of magnitude within the deflagration front  due to electron heat conduction increasing strongly with temperature \cite{Modestov-09}.

As explained in Sec. IV C, the smooth relaxation tail in the doped
region is due to high mobility of holes in that region and the Nernst
term modeling the injection barrier. The numerical solution shown in
Fig. 3 uses the parameter value $f = 1.1$, which provides the
characteristic length scale  $ \approx 71L_{p} $ of the relaxation. Indeed,
in Fig. 3 we see  that the relaxation length scale in the doped region
exceeds the length scale of ion diffusion $L_{p} $ by approximately  two
orders of magnitude. Figure 4 shows modifications of the front structure
caused by changing of the parameter $f$. In agreement with the theoretical
predictions, relaxation to the final doped state becomes smoother as $f$
approaches unity. At the same time, the parameter $f$ does not influence the head of the front.

\begin{figure}
\includegraphics[width=3.6in,height=2.8in]{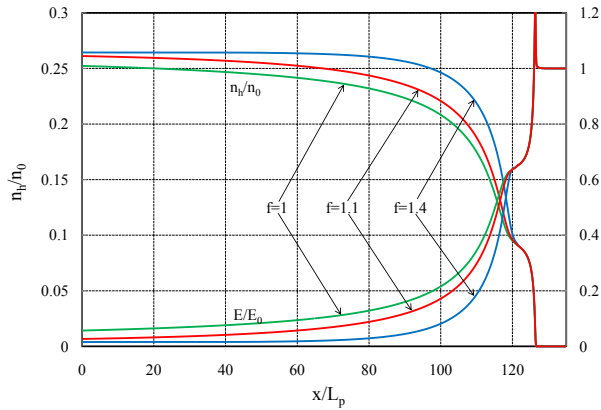}
\caption{Concentration
 of holes and the electric field inside the p-doping front
for different values of the $f$-factor.}
\end{figure}

The details of the p-front structure in the undoped region (head of
the front) are shown in Fig. 5. Figure 5(a) presents the concentration
of holes and the electric field. In agreement with the asymptotic theory
of Sec. IV D, we can see sharp gradients of the hole concentration
with characteristic length scales noticeably smaller than the length
scale $L_{p} $ of ion diffusion.
\begin{figure}
\subfigure[]{\includegraphics[width=3.6in,height=2.8in]{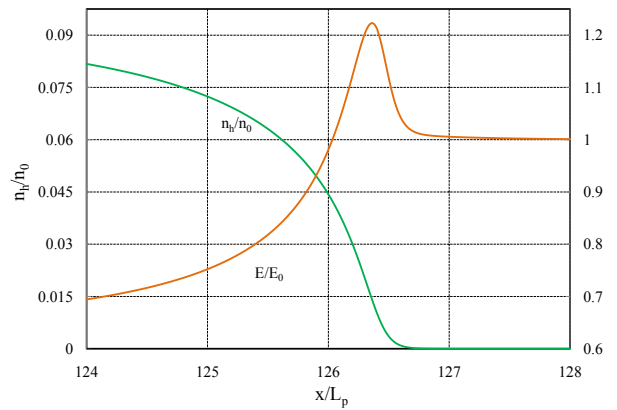}}
\subfigure[]{\includegraphics[width=3.6in,height=2.8in]{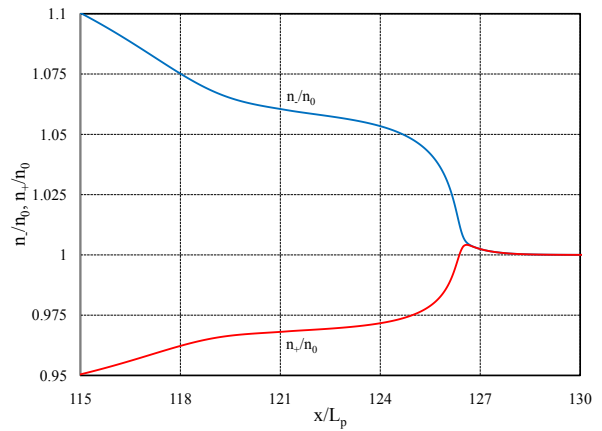}}
\caption{Structure of the p-front in the undoped region, (a) hole
concentration and the electric field, (b) ion concentrations.}
\end{figure}
The theory predicts that the electric field in the leading part of the
front is proportional to the spatial derivative of the hole
concentration, see Eq. (\ref{eq70}), which leads to a sharp peak in
the electric field clearly visible in Fig. 5(a). Figure 5(b) shows
the variations of the ion concentration at the head of the doping front.
Unlike the hole concentration, the concentrations of ions demonstrate a more
complicated behavior. First, the two  ion concentrations approach each other
on a short length scale related to the mode with the eigenvalue
given by Eq. (\ref{eq69}). However, instead of a monotonic
relaxation to the limiting values, the concentration $n_{ +}  $  overshoots
$n_{0} $, so that $n_{ -}  $ and $n_{ +}  $ meet at some value
exceeding $n_{0} $ (in agreement with the constraint of quasi-neutrality). After that, both the ion concentrations $n_{ -}  $
and $n_{ +}  $ approach the limiting value $n_{0} $ together from
above. Such a specific way of relaxation for the ion concentrations is in
agreement with the second mode, as given by Eq. (\ref{eq72}), predicted
analytically in Sec. IV D. The ion relaxation is of course determined by ion
diffusion and occurs on  length scales comparable to $L_{p} $.

The specific behavior of the doping front parameters close to the critical point
is another interesting feature of the system. The critical
point was not discussed in Ref. \cite{Modestov-10}, since in that work a matching of the
concentration gradients was done by a linear extrapolation from the doped and undoped
regions. However, we stress that the critical point is \textit{not} a mathematical artifact
of the phenomenological transition function $\psi $. As demonstrated in
Sec. IV E, the local velocity of holes produced by their mobility is in resonance
with the p-front velocity in the critical point, and this resonance explains the physical
origin of the critical point and the zone around it. In order to understand the
effect of this critical zone better, we may integrate Eqs. (\ref{eq45})-(\ref{eq50})
in two opposite directions: from the doped region with $\psi = 1$ (from left
to right in Fig. 3) and from the undoped region with $\psi = 0$ (from right
to left in Fig. 3).
\begin{figure}
\includegraphics[width=3.6in,height=2.8in]{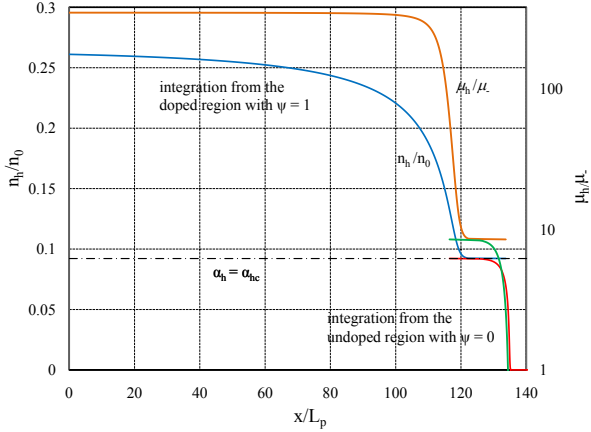}
\caption{Hole concentration and mobility inside the p-doping front obtained by integration from the doped region (left) with the Nernst term and from the undoped region (right) without the Nernst term.}
\end{figure}
Figure 6 presents the concentration of holes and the hole mobility obtained
from such an integration procedure. As illustrated in Fig. 6, integrating from the doped
region we could not reach the undoped one, and vice versa. Instead,
using both ways of integration we reach asymptotically  the
critical point as a saturation limit: from above for $\psi = 1$ and
from below for $\psi = 0$. From the mathematical point of view,
passing the critical resonance point means changing sign of the
right-hand side of Eq. (\ref{eq45}), $\left( {CM_{h} \varepsilon -
1} \right)$. The continuous transition from the doped (left) to
undoped (right) regions requires a non-positive
derivative of the hole concentration, $d\alpha _{h} / d\xi $, and,
therefore, the effective hole diffusion $\gamma _{eff} $ has to
change sign in the critical point. We remind that the effective
diffusion is a combination of the real diffusion and the Nernst contribution
describing the injection barrier. The role of the Nernst term is
controlled by the phenomenological function $\psi $, which therefore
has to change from 1 (the Nernst term is then switched on) to 0 (the Nernst
term is switched off) in the vicinity of the critical point. For example,
matching the concentration profiles obtained by integration from
the right and from the left at the critical point we find the front
structure for $\psi (n_{h} )$ in the form of a step-function.
Still, a
smooth transition function is  required for treating   the doping
fronts numerically within the  evolution problems. Due to the lack of a good
thermodynamic model for the transition from the undoped to doped state,  we use the following phenomenological form of the
transition function:
\begin{eqnarray}
\label{eq78}
&&\!\!\!\!\!\! \psi = 0.5 + 0.5\tanh {\left[ {A_{1} \left(
{CM_{h} \varepsilon - 1} \right){\left| {CM_{h} \varepsilon - 1}
\right|}^{n}} \right.}
\nonumber \\ &&
{\left. {\quad\quad \quad \quad \quad \quad + 0.5 \ln \left(
\alpha _{h\infty}/ \alpha _{h} - 1 \right)} \right]}.
\end{eqnarray}
The hyperbolic tangent of Eq. (\ref{eq78}) provides a smooth transition from 1 to 0
for $\psi $ as we go from the doped to undoped region. The function Eq.
(\ref{eq78}) depends on the combination $\left( {C\mu _{h} \varepsilon - 1}
\right)$, which is the right side of Eq. (\ref{eq45}). When this combination is zero,
the residual term makes $\gamma _{eff} = 0$ and allows for a smooth change in
all the concentrations. The other parameters of the transition function are
chosen to reduce the plateau near the critical point. In our calculations presented in Fig. 3 we used $A_{1} = 8$ and $n
= 0.1$. Hypothetically, a transition function $\psi $ may exist, which
eliminates the plateau completely. However, we believe that it is important to obtain the transition function from first principles of
thermodynamics and quantum mechanics, rather than to make a more elaborate phenomenological
construction. This is indeed an important and difficult problem left for the
future, while at present we simply use a phenomenological function for $\psi
$ to obtain a smooth transition through the critical point from the undoped to
doped region.

All the main characteristic features of the doping front for holes are relevant for
electron doping as well. In Fig. 7 we depict the stationary doping front
for electrons (the n-front). The characteristic length scale related to the
n-doping front is defined in the same way as for the p-front according to
\begin{figure}
\centering
\includegraphics[width=3.6in,height=2.8in]{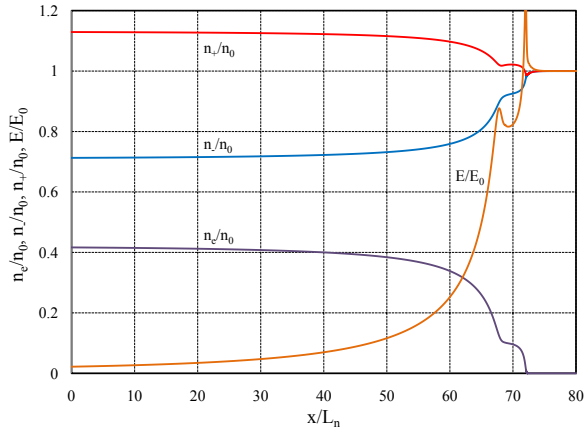}
\caption{Concentrations of electrons and ions, and the electric field inside the n-doping front.}
\end{figure}
\begin{equation}
\label{eq79}
L_{n} = {\frac{{D_{ -}} } {{U_{n0}}} } = {\frac{{D_{ -}  n_{1e}}} {{(\mu _{
+}  + \mu _{ -}  )n_{0} E_{0}}} }.
\end{equation}
In order to obtain a numerical result for the front structure we used the electron mobility
function
\begin{equation}
\label{eq80}
M_{h} = 145{\left[ {1 + \tanh \left( {21.6\alpha _{h} - 4.3} \right)}
\right]}
\end{equation}
constructed from the experimental data
\cite{Arkhipov,Shimotani}, and the phenomenological transition
function similar to the one given in Eq. (\ref{eq78}), i.e.
\begin{eqnarray}
\label{eq81}
&&\!\!\!\!\!\! \psi _{e} = 0.5 + 0.5\tanh {\left[ {14.4\left(
{CM_{e} \varepsilon - 1} \right){\left| {CM_{e} \varepsilon - 1}
\right|}^{0.1}} \right.}
\nonumber \\ &&
{\left. {\quad \quad \quad \quad \quad \quad \quad + 0.5\ln \left(
\alpha _{e\infty}/ \alpha _{e} - 1 \right)} \right]}.
\end{eqnarray}
 The electron mobility in
OSCs is somewhat lower than the hole mobility.
Similar to the p-front shown in Fig. 3, the n-doping fronts have
also very elongated tails in the doped region, a sharp head
with a peak of the electric field in the undoped region and a
critical point of transition from the undoped to doped state at $\mu
_{e} / \mu _{ -}  \approx 3.5$. We point out that the critical
point for electrons corresponds to considerably lower mobility (the critical point for holes is achieved at $\mu _{h} /
\mu _{ -}  \approx 8.5)$. Because of the lower electron mobility,
the critical point demonstrates a more complicated structure for the
n-front in comparison with the p-front. In the case of the p-front,
saturation of all the concentrations and the electric field to the
plateau of the critical point occurs monotonically, see Fig. 6. On
the contrary, in the case of the n-doping front presented in Fig. 7,
the concentration of positive ions and the electric field exhibit a
non-monotonic behavior when approaching the plateau. This behavior
shows clearly existence of two perturbation modes in the doped zone
close to the critical point, which resemble qualitatively the modes
obtained in Sec. IV D. We have found a qualitatively similar
structure of the transition zone for several types of function $\psi
$.
%At the same time, we cannot exclude existence of a function $\psi $, for which only one mode dominates in the approach to the plateau similar to the p-front shown in Figs. 3, 6. Still, similar to holes, the transition function should be obtained from the basic principles of thermodynamics.

An interesting consequence of the above discussion of the critical point
is the possibility of a weak doping process, when the final concentration of
the light charge carrier is still smaller than the corresponding critical value. Such
a doping front can be described without the Nernst potential and the
transitional function. At the same time, the front structure in the undoped region  remains the same as described in Sec.
IV D. The p-front of such a weak doping transformation is shown in Fig. 8;
it is similar to the right part (head) of the front in  Figs. 3 and 6.
\begin{figure}
\centering
\includegraphics[width=3.6in,height=2.8in]{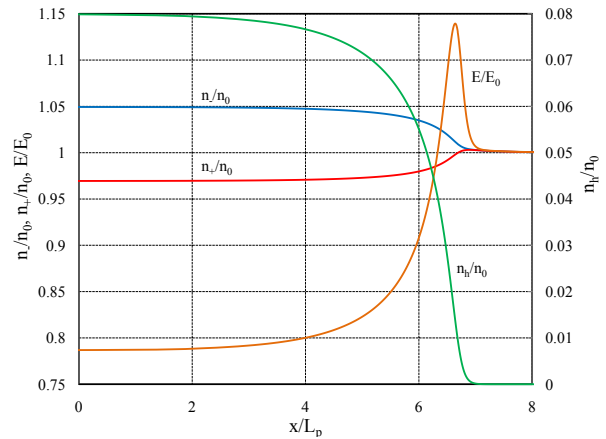}
\caption{Concentrations of holes and ions, and the electric field inside the p-front of weak doping for  $n _{1h}/n _{0} = 0.08$.}
\end{figure}
The final hole concentration in Fig. 8 is smaller than the critical one,
namely, $\alpha _{1h} = 0.08$, and the corresponding final mobility of the holes is
$\mu _{1h}/\mu _{0} = 4.5$. As the hole mobility is about 80 times smaller
than in fully doped case, then, according Eq. (\ref{eq77}), the final electric
field remains relatively large behind the front, $\varepsilon _{1} \approx 0.8$, though smaller than the initial one, $\varepsilon _{0} = 1$. The
total length of the front is much shorter than the front width in Fig. 3, as
the long relaxation tail in the strongly doped region  is missing here. In general, the front of weak doping transformation may be interpreted as a part of the complete doping front from the head to the critical point.

Finally, we present the structure of the p- and n-doping fronts as they
accelerate towards each other in Fig. 9. Since the characteristic time scales
related to the fronts (that is $\tau _{p} \propto L_{p} / U_{p} = D_{ -}  /
U_{p}^{2} $ and $\tau _{n} \propto L_{n} / U_{n} = D_{ -}  / U_{n}^{2} )$
are much smaller than the time of the front acceleration, then the structure may
be obtained within the quasi-classical Wentzel-Kramers-Brillouin
approximation. Within this approximation, structure of the fronts remains
self-similar, but the length scales $L_{p} $ and $L_{n} $ decrease due to
the increase of the electric field and the front velocities $U_{p} $ and
$U_{n} $ as discussed in Sec. III.

\begin{figure}
\includegraphics[width=3.6in,height=3.4in]{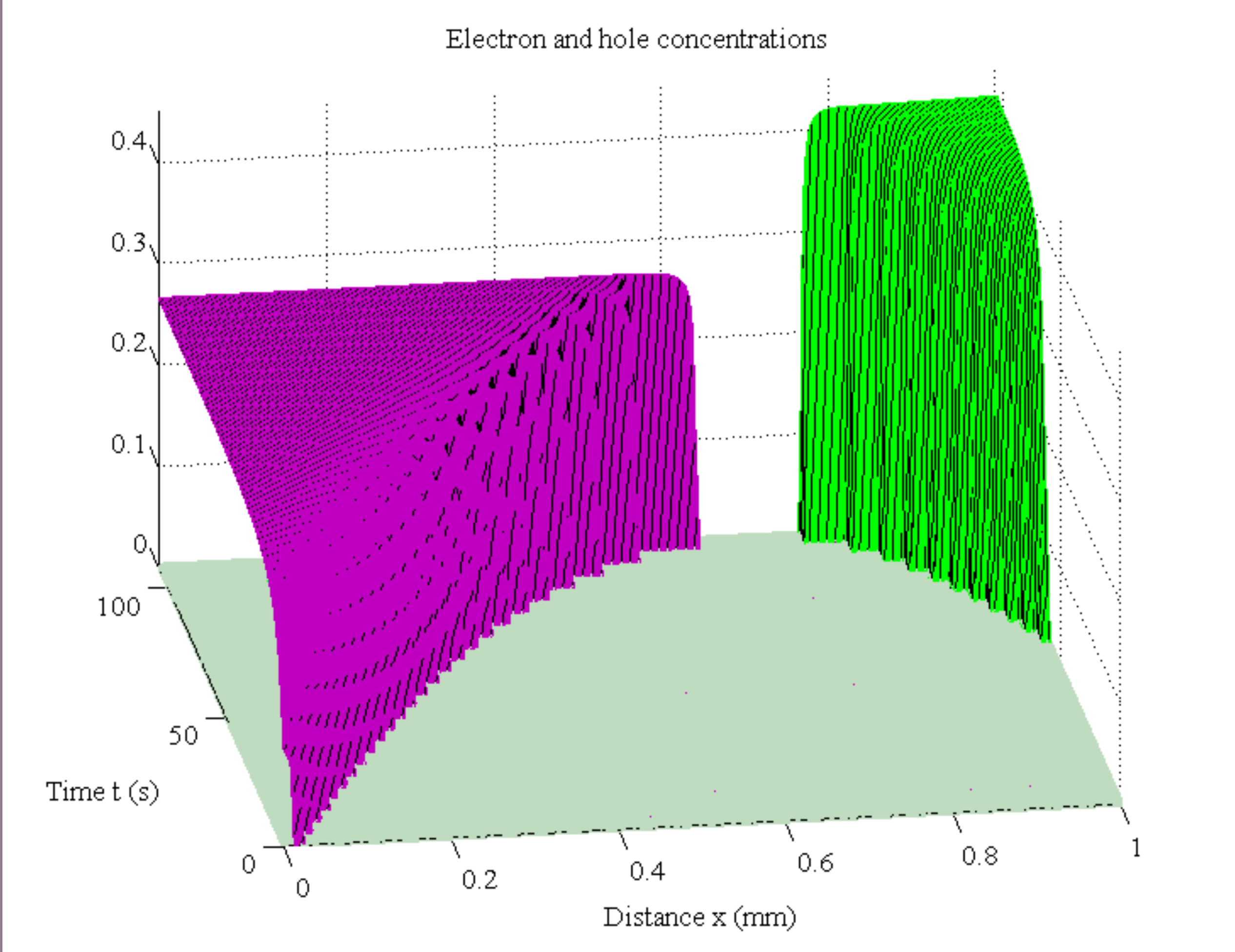}
\includegraphics[width=3.6in,height=3.4in]{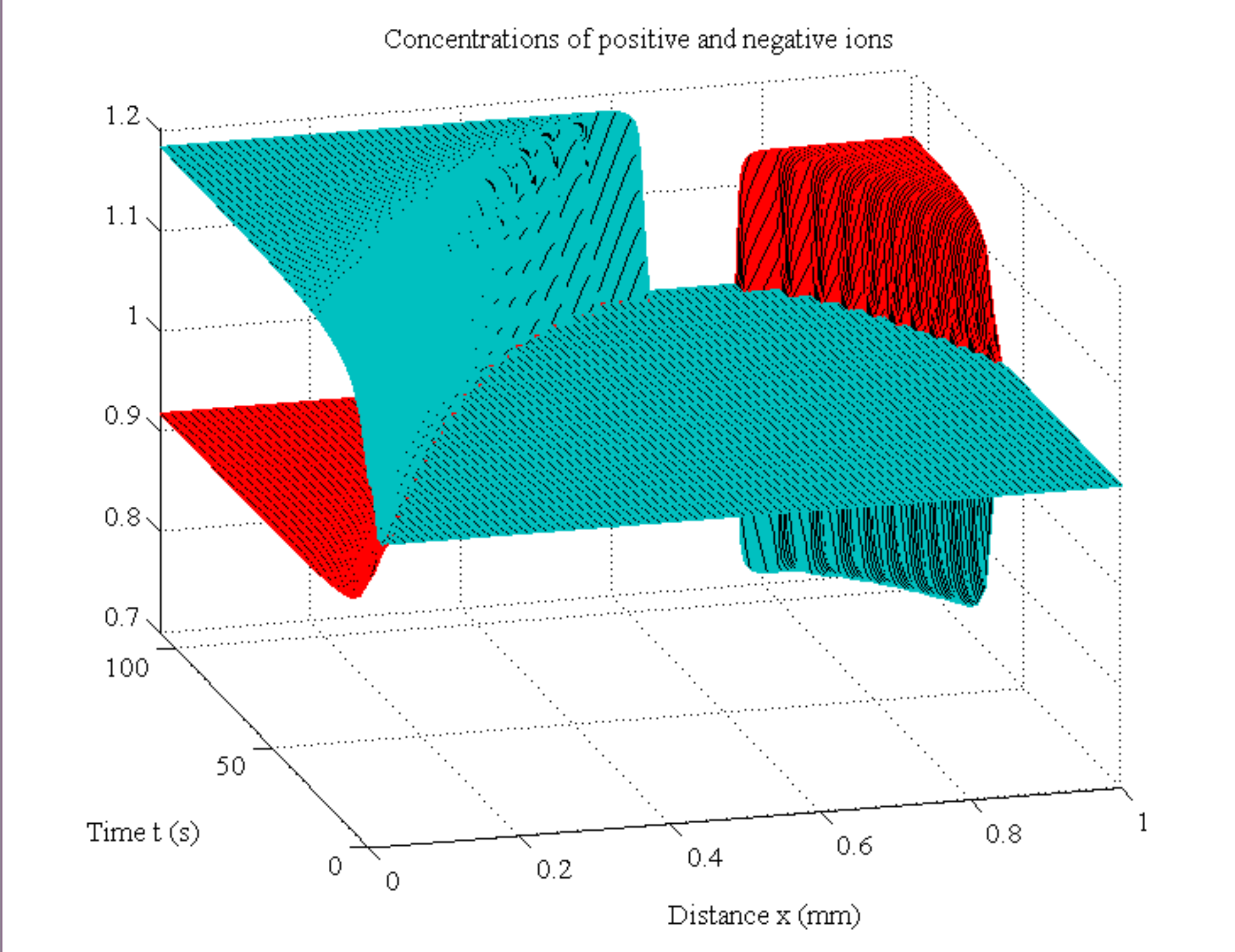}
\caption{The p- and n- fronts accelerating towards each other. (a) hole
and electron concentrations are depicted by magenta and green, (b) negative and positive ions are shown by blue and red. }
\end{figure}

\section{Conclusions}

In this paper we have investigated the dynamics and internal structure
of the planar p- and n-type doping fronts in organic semiconductors with
applications to LECs. The study is based on the drift-diffusion
model taking into account the  injection barrier and a strongly
nonlinear dependence of the hole and electron mobility/diffusion
on concentration, following the work  Ref. \onlinecite{Modestov-10}. A similar
model has been employed before to describe transformation fronts in
one-electrode devices like electrochemical sensors and actuators in Ref.
\onlinecite{Wang-09}. Still, there is an important difference between
these two types of processes/devices. In the doping front (studied
in the present paper) holes and electrons populate the active
material while ions give way to the light charges. As a result,
conductivity of the material \textit{increases} drastically, by 2-3 orders of
magnitude. On the contrary, the transformation fronts in ambipolar
devices imply that cathions replace holes with strong \textit{decrease} of
conductivity. The different characters of the processes lead naturally to
different properties of the transformation fronts.

Here we have studied parameters of the doping fronts on different
scales, both within the discontinuous front approach and also by taking into
account the internal front structure. Within the limit of a
discontinuous front we have derived the analytical formulas for the front
velocities, the ion concentrations in the doped region, and described
dynamics of the p- and n-doping fronts accelerating towards each
other in LECs. The analytical results for the planar front
dynamics are in a good quantitative agreement with the experimental
data for the slowest part of the experimentally observed
front brush. We remind that experiments demonstrate also a
complicated multidimensional front dynamics related to
instabilities. Theoretical investigation of the front instabilities
is beyond the scope of this paper and is presented elsewhere
\cite{Bychkov-sub}.

One of the main purposes of the present paper was to investigate the
internal structure of the doping fronts. In agreement with the
previous ideas \cite{Modestov-10}, we show that continuous
transition from the doped to undoped state in the form of a moving
front requires a thermodynamic injection barrier and a nonlinear
dependence of the hole/electron mobility on concentration, which is
quite in line with the basic principles of the doping process. We
have studied the  asymptotic behavior of the front parameters: 1) relaxation to the doped state at the
back of the front, 2) deviation from the undoped state at the head
of the front and 3) the critical point of the transition from the doped
to undoped parts of the front, where the velocity of the light charges
are in resonance with the front speed. We have also obtained a numerical
solution for the front structure. Both the analytical theory and the
numerical solution demonstrated the multi-scale features of the
doping fronts, which include an extremely smooth relaxation tail
in the doped region, a sharp head of the front with large
gradients in the undoped zone, and a plateau at the critical point.
The described front structure agrees qualitatively with observations
of the previous experiments \cite{Modestov-10}.
% still, in order to perform a quantitative comparison a refined experimental data is needed, which is not available at present.

\bigskip

\acknowledgements

The authors are grateful to Ludvig Edman and Piotr Matyba for numerous discussions and the experimental data. This  work was supported by the Swedish Research Council (VR) and by
the Kempe Foundation.

\bigskip

%\newpage


\begin{thebibliography}
\bibliographystyle{}

\bibitem{Malliaras} G. Malliaras, R. Friend, Phys. Today \textbf{58}(5), 53 (2005).

\bibitem{Sirringhaus} H. Sirringhaus, N. Tessler, R. N. Friend, Science \textbf{280}, 1741 (1998).

\bibitem{Heeger} A. J. Heeger, Rev. Mod. Phys. \textbf{73}, 681 (2001).

\bibitem{Forrest} S. R. Forrest, Nature \textbf{428}, 911 (2004).

\bibitem{Leger} M. J. Leger, Adv. Mater. \textbf{20}, 837841 (2008)

\bibitem{Chiang} C. K. Chiang, C. R. Fincher, Y. W. Park, A. J. Heeger, H. Shirakawa, E. J. Louis, S. C. Gau, and A. G. Macdiarmid, Physical Review Letters \textbf{39}, 1098 (1977).

\bibitem{Li} Y. Li, Y. Cao, J. Gao, D. Wang, G. Yu, A. J. Heeger, Electrochemical
Synthetic Metals \textbf{99}, 243 (1999).

\bibitem{Pei-96} Q. B. Pei, Y. Yang, G. Yu, C. Zhang, and A. J. Heeger,
Journal of the American Chemical Society \textbf{118}, 16, 3922
(1996).

\bibitem{Matyba-08} P. Matyba, M. R. Andersson, L. Edman, Organic Electronics, \textbf{9}, 699 (2008).

\bibitem{Pei-95} Q. B. Pei, G. Yu, C. Zhang, Y. Yang, A. J. Heeger, Science \textbf{269}, 1086 (1995).


\bibitem{Coropceanu} V. Coropceanu, J. Cornil, D. A. da Silva, Y. Olivier, R. Silbey, and J. L. Bredas
Chem. Rev. \textbf{107}, 926 (2007).

\bibitem{Bredas} J. L. Bredas, J. P. Calbert, D. A. da Silva, J. Cornil, PNAS \textbf{99} (9), 5804 (2002).

\bibitem{Arkhipov} V. I. Arkhipov, E. V. Emalianova, P. Heremans, H. Bassler
Phys. Rev. B \textbf{71}, 235202 (2005).

\bibitem{Pei-97} Q. B. Pei, Y. Yang, G. Yu, Y. Cao, and A. J. Heeger, Synthetic Metals \textbf{85}, 1229 (1997).

\bibitem{Sun} Q. J. Sun, Y. F. Li, and Q. B. Pei, Journal of Display Technology \textbf{3}, 211 (2007).

\bibitem{Matyba-09} P. Matyba, K. Maturova, M. Kemerink, N. D. Robinson, L. Edman, Nature Mater. \textbf{8}, 672 (2009).

\bibitem{Reenen} S. van Reenen, P. Matyba, A. Dzwilewski, R. A. J. Janssen, L. Edman, and M. Kemerink, J. Am. Chem. Soc. \textbf{132}, 13776 (2010).

\bibitem{Gao-04} J. Gao and J. Dane, Appl. Phys. Lett. \textbf{84}, 2778 (2004).

\bibitem{Hu-06} Y. F. Hu, C. Tracy, and J. Gao, Appl. Phys. Lett. \textbf{88} (2006).

%\bibitem{Gao} J. Gao, J. Dane, Appl. Phys. Lett. \textbf{84}, 15, 2778 (2004).

%\bibitem{Matyba-10} P. Matyba, H. Yamaguchi, G. Eda, M. Chhowalla, L. Edman, and N. D. Robinson, ACS Nano, \textbf{4}, 637 (2010).

%\bibitem{Yu} Z. Yu, L. Hu, Z. Liu, M. Sun, M. Wang, G. Gruner, and Q. Pei, Appl. Phys. Lett. \textbf{95}, 3304 (2009).

%\bibitem{Rodovsky} D. B. Rodovsky, O. G. Reid, L. S. C. Pingree, and D. S. Ginger, ACS Nano \textbf{4} (5), 2673 (2010).

%\bibitem{Slinker} J. D. Slinker, J. A. DeFranco, M. J. Jaquith, W. R. Silveira, Y. W. Zhong, J. M. Moran-Mirabal, H. G. Craighead, H. D. Abruna, J. A. Marohn, G. G. Malliaras, Nature Mater. \textbf{6}, 894 (2007).

\bibitem{Johansson} T. Johansson, N. K. Persson, O. Inganas, J. Electrochem. Soc. \textbf{151}, E119 (2004).

\bibitem{Robinson} N. Robinson, J. H. Shin, M. Berggren M., L. Edman, Phys. Rev. B \textbf{74}, 155210 (2006).

\bibitem{Smith} D. L. Smith, J. Appl. Phys. \textbf{81}, 2869 (1997).

\bibitem{Manzanares} J. A. Manzanares, H. Reiss, A. J. Heeger, J. Phys. Chem. B \textbf{102}, 4327 (1998).

\bibitem{Lacroix} J.C. Lacroix, K. Fraoua, P.C. Lacaze, J. Electroan. Chem. \textbf{444}, 83 (1998).

\bibitem{Miomandre} F. Miomandre, M.N. Bussac, E. Vieil, L. Zuppiroli, Chem. Phys. \textbf{255}, 291 (2000).

\bibitem{Wang-04} X. Wang, B. Shapiro, E. Smela, Adv. Mater. \textbf{16}, 1605 (2004).

\bibitem{Wang-09} X. Wang, B. Shapiro, E. Smela, J. Phys. Chem. C \textbf{113}, 382 (2009).

\bibitem{Modestov-10} M. Modestov, V. Bychkov, G. Brodin, D. Valiev, M. Marklund, P. Matyba,
L. Edman, Phys. Rev. B \textbf{81}, 081203(R) (2010).

\bibitem{Bychkov-sub} V. Bychkov, P. Matyba, V. Akkerman, M. Modestov, D. Valiev, G. Brodin,
C.K. Law, M. Marklund, L. Edman, \textit{submitted}.

\bibitem{Shimotani} H. Shimotani, G. Diguet, and Y. Iwasa, Appl. Phys. Lett. \textbf{86}, 022104 (2005).

\bibitem{Modestov-09} M. Modestov, V. Bychkov, D. Valiev, M. Marklund, Phys. Rev. E \textbf{80} 046403 (2009).

\end{thebibliography}
\end{document}